\documentclass{emulateapj}
\usepackage{epsfig,natbib}
\def\lap{\lower.5ex\hbox{$\; \buildrel < \over \sim \;$}}
\def\gap{\lower.5ex\hbox{$\; \buildrel > \over \sim \;$}}

\begin{document}

\title{The ACS Nearby Galaxy Survey Treasury XI. The Remarkably Undisturbed NGC 2403 Disk}

\author{Benjamin F. Williams\altaffilmark{1},
Julianne J. Dalcanton\altaffilmark{1},
Adrienne Stilp\altaffilmark{1},
Andrew Dolphin\altaffilmark{2},
Evan D. Skillman\altaffilmark{3},
David Radburn-Smith\altaffilmark{1}
}
\altaffiltext{1}{Department of Astronomy, Box 351580, University of Washington, Seattle, WA 98195; ben@astro.washington.edu; jd@astro.washington.edu;adrienne@astro.washington.edu}
\altaffiltext{2}{Raytheon, 1151 E. Hermans Road, Tucson, AZ 85706; dolphin@raytheon.com}
\altaffiltext{3}{Department of Astronomy, University of Minnesota, 116 Church
St. SE, Minneapolis, MN 55455; skillman@astro.umn.edu}


\begin{abstract}
We present detailed analysis of color-magnitude diagrams of NGC~2403,
obtained from a deep (m$\lap$28) Hubble Space Telescope (HST) Wide
Field Planetary Camera 2 observation of the outer disk of NGC~2403,
supplemented by several shallow (m$\lap$26) HST Advanced Camera for
Surveys fields. We derive the spatially resolved star formation
history of NGC~2403 out to 11 disk scale lengths.  In the inner
portions of the galaxy, we compare the recent star formation rates (SFRs) we
derive from the resolved stars with those measured using {\it GALEX}
FUV + {\it Spitzer} 24$\mu$ fluxes, finding excellent agreement
between the methods.  Our measurements also show that the radial gradient in recent SFR mirrors
the disk exponential profile to 11 scale lengths with no break,
extending to SFR densities a factor of $\sim$100 lower
than those that can be measured with {\it GALEX} and {\it Spitzer}
(${\sim}2{\times}10^{-6}$~M$_{\odot}$~yr$^{-1}$~kpc$^{-2}$).  Furthermore, we find that the cumulative
stellar mass of the disk was formed at similar times at all radii.    We
compare these characteristics of NGC~2403 to those of its
``morphological twins,'' NGC~300 and M~33, showing that the structure and age
distributions of the NGC~2403 disk are more similar to those of the
relatively isolated system NGC~300 than to those of the Local Group
analog M~33. We also discuss the environments and H{\sc i} morphologies of
these three nearby galaxies, comparing them to integrated light
studies of larger samples of more distant galaxy disks.  Taken
together, the physical properties and evolutionary history of NGC~2403 suggest that the galaxy has had no close encounters
with other M~81 group members and may be falling into the group for the
first time.

\end{abstract}

\section{Introduction}

Most of the star formation occurring in the present epoch takes place
in galaxy disks \citep[e.g.,][]{williams2011}, but most disks are
difficult to study in detail due to the complexities of their bulges
\citep[e.g.,][]{byun1995}.  However, some spirals are nearly
bulgeless, making them excellent laboratories for the study of star
formation and disk growth in relatively undisturbed disks.  While
theoretical models suggest that post-merger galaxies can have disk
fractions of up to $\sim$90\% \citep{hopkins2009a,hopkins2009b}, the
delicate construction of these bulgeless galaxies is consistent with
long-term evolution that lacks strong recent tidal interactions or
mergers, which tend to disturb such structure \citep{barnes1992}.

While bulgeless disks appear to be ideal for studying undisturbed disk
evolution, most are too distant to study in great detail.  While
integrated color or spectral profiles
\citep[e.g.,][]{macarthur2003,yoachim2012} provide large samples and
have yielded great insight, significant additional knowledge can be
gained by studying nearby systems.  In the nearby universe
(D$<$4~Mpc), there are three nearly identical bulgeless systems (M~33, NGC~300 and NGC~2403) whose
stellar populations can be resolved well below the tip of the red
giant branch (TRGB).  With resolved photometry of
these systems, we can compare the detailed evolution of these
star-forming disks.  Such comparative studies offer the opportunity to
search for clues to the processes which drive the secular evolution of
galaxy disks.

NGC~2403 is a SAB(s)cd galaxy \citep{devaucouleurs1991} with an
inclination of 63$^{\circ}$ \citep{fraternali2002}, position angle of
124$^{\circ}$ \citep{deblok2008}, a distance modulus of 27.5
\citep[3.2
Mpc,][]{tammann1968,freedman1988,dalcanton2009,radburn2011}, and scale
length of 2~kpc (\citealp{cepa1988,barker2012}, hereafter B12).  It
lies on the outskirts of the M~81 group of galaxies, more than a Mpc
away (projected distance) from M~81. 

\begin{figure}
\centerline{\psfig{file=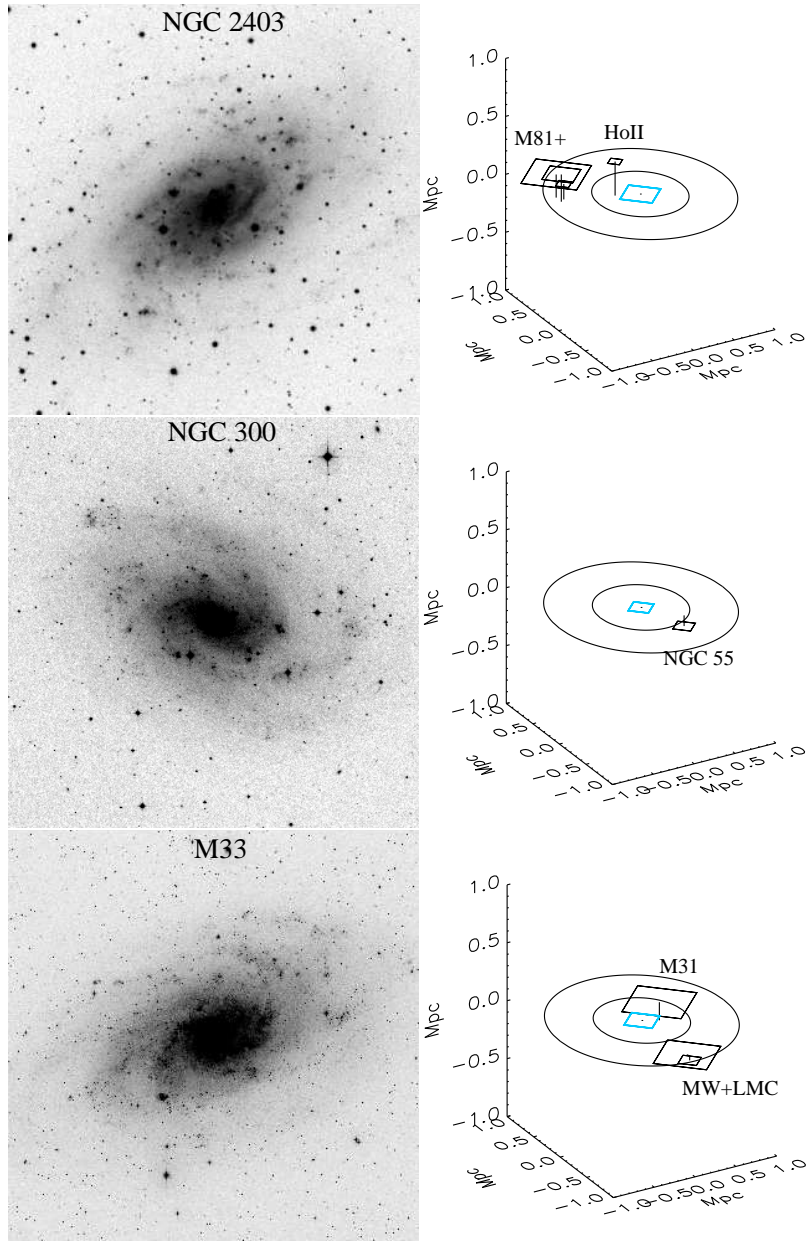,height=6.5in,angle=0}}
\caption{Comparison of the 3 Local Volume pure-disk galaxies.  Left
column shows images of 15$\times$15~kpc digitized sky survey images of
each.  NGC300 has been flipped about the North-South (up-down) axis
in order to show the arms bending the same direction as the other two
for an easier comparison.  Right column shows schematics of the
environments of the galaxies, with all galaxies brighter than
$M_B=-16$ shown.  Larger symbols denote brighter (more massive)
galaxies. }
\label{compenvs}
\end{figure}

In spite of their similarities to NGC~2403 in morphology, metallicity,
star formation rate (SFR), gas content, and luminosity (summarized in
Table~\ref{summary}), M~33 and NGC~300 display some interesting
differences.  The data in the table show that NGC~300 is a somewhat scaled-down version of the other
two, with a smaller scale length and lower luminosity.  However, the
most noticeable differences are in their environments, which we show
in Figure~\ref{compenvs}.  M~33 is relatively close to M~31, with which
it has likely interacted in the past
\citep[e.g.,][]{mcconnachie2010,bernard2012}; M~33 also shows evidence
for recent weak tidal interactions \citep[tidal index =
2.0,][]{karachentsev2004}, consistent with its warped H{\sc i} disk
\citep{rogstad76, deul87, corbelli97}.  NGC~300 is in the foreground of
the Sculptor Group \citep{karachentsev03}, and is relatively isolated
compared to M~33.
%, with a tidal index of $-$0.2.  
Still, it does form a gravitationally bound pair with the dIrr galaxy
NGC~55 \citep{tully06} which is comparable in luminosity.  In addition,
its H{\sc i} disk shows a severe warp in its outer parts
\citep{puche90} which may be due to the proximity of NGC~55.  Of the
three morphologically analogous galaxies, NGC~2403 is the most
isolated as it is an outlying member of the M~81 group, with no nearby
massive neighbors.  The H{\sc i} distribution and velocity field of
NGC~2403 is extremely symmetric \citep{fraternali2001,fraternali2002,deblok2008}, showing no evidence of recent
interactions. Herein, we explore the
possibility that these environmental differences may be responsible for
differences not only in their disk structures, but also in their detailed
formation histories.

The properties of the three nearby bulgeless systems suggest that
local environment may play an important role in the formation of disk
breaks, although the role of environment in the formation and evolution of disk
breaks is a complex problem.  Empirical studies of breaks in field and
cluster samples find no differences between the two
\citep{maltby2012,erwin2012}.  If the environment is responsible for
disk breaks, it is unclear why there would be many unbroken disks in
galaxy clusters. However, comparing the cluster environment to the
field environment may not be as sensitive as comparing the local
environments of individual field galaxies.  In our sample, the disk
with the closest massive companion (M~33) has a clear break, while
those without a massive companion do not show such a break
(\citealp{bland2005,ferguson2007}; B12). Thus,
among these three field galaxies, local interactions may play a
significant role in the production of breaks.  The influence of a
large cluster may still work to smooth over disk breaks.  The cluster
environment clearly has profound effects on the evolution of disks, as
ram pressure stripping may prematurely truncate the outer disk gas
supply, inducing late-type disks to transform into S0 disks.

At the same time, the effects of radial migration of stars due to
internal dynamics can significantly affect the production of disk
breaks \citep{roskar2008}.  The lack of disk breaks in NGC~2403 and
NGC~300 is not a sensitive indicator of the presence or absence of
radial migration.  As seen in the simulations presented by
\citet[][see their Figure 2]{radburn-smith2012}, dynamical effects
such as radial migration together with ongoing star formation in the
outer disk can act to smooth out an initial disk break on short
timescales. Thus we cannot distinguish between a slight break that has
been quickly smoothed over by radial migration, or no break with
little radial migration.  This intricate interplay of processes is
also explored in the recent study of \citet{roediger2012}, who show
that no single model of radial migration, gas accretion, or warping
can account for the radial distribution of stellar populations and
luminosity in the disk galaxies in Virgo. Thus, the current picture is
complex, with no single dominant process responsible for the
appearance of the disk break, or its subsequent evolution.  Our
comparison of local disks will add some detailed constraints to this
difficult problem.

Furthermore, there is a possible environmental dependence on the
relationship between the gas disk structure and disk breaks within our
sample.  Studies of larger samples of disks show evidence that breaks
are associated with H{\sc i} warps. Edge-on disks that show breaks
also tend to have warped H{\sc i} disks, whereas those without breaks
tend to have undisturbed H{\sc i} disks \citep{vanderkruit2007}.
Therefore, the formation of H{\sc i} warps and stellar breaks may be
related.  Specifically, they could both be the result of the outer
disk forming from late infall of high angular momentum gas or the
result of
gravitational interactions with other galaxies
\citep{vanderkruit2007}.  The most isolated pure disk in the Local
Volume (NGC~2403) shows no break and no warp in H{\sc i}.  The disk with a
single relatively distant low-mass companion (NGC~300) shows no break,
but has a warped H{\sc i} disk.  The disk with the most nearby
massive companions has a clear break and a warped H{\sc i} disk.
This sequence is consistent with disk breaks being induced by tidal
interactions that are strong enough to strip outer gas and/or funnel
it to smaller radii. It will be of interest to check the agreement of
this sequence with a larger sample when more pure disks have detailed
constraints on their evolution and environment, including precise TRGB
distances and reliable star formation histories (SFHs), as the galaxies
included here.

Finally, differences in the evolutionary history of the disks show
hints of environmental effects.  Detailed studies of M~33 and NGC~300
reveal significant differences in SFH derived
from resolved stellar photometry
\citep{barker2007,williams2009b,barker2011,gogarten2010}.  In M~33, the
stellar populations show an age inversion at the M~33 disk break
\citep{williams2009b}, while the NGC~300 populations show no
significant trends with radius \citep{gogarten2010}.  Existing studies
of NGC~2403 hint that its history may be more similar to that of NGC~300
than to that of M~33, possibly indicating a mostly isolated past.
Recent stellar populations studies suggest that star formation is
present out to at least 6 scale lengths \citep{davidge2007}, but these
studies have been mainly from the ground
(\citealp{davidge2002,davidge2003,davidge2007}; B12), although B12
augment their ground-based data with a few ACS archival fields.  B12
found no evidence for a metallicity gradient in the old stellar
population.  This detailed study also found evidence for a very
extended stellar component in the far outer disk, beginning at $\sim$8
scale lengths.

Herein, we focus on {\it HST} observations to probe the disk evolution
of NGC~2403 in spatial and temporal detail similar to {\it HST} studies
of M~33 \citep{barker2007,williams2009b} and NGC~300
\citep{gogarten2010}. Recently, the ACS Nearby Galaxy Survey Treasury
\citep[ANGST,][]{dalcanton2009}, and the GHOSTS Surveys
\citep{radburn2011} obtained significant amounts of HST imaging of the
outer disk of NGC~2403.  These observations, along with archival
imaging of the inner disk, provide the deepest resolved stellar
photometry ever obtained in NGC~2403. We analyze all of these {\it HST}
data sets homogeneously to measure the SFH as a
function of radius in NGC~2403. With this new information, we can now
compare the structural evolution of NGC~2403, NGC~300, and M~33, and
explore the role that environment may have played in producing their
differences. We assume the distance, inclination angle, and position
angle reported above for all conversions between location and
galactocentric radius.

\section{Data Acquisition and Reduction}

\begin{figure}
\centerline{\psfig{file=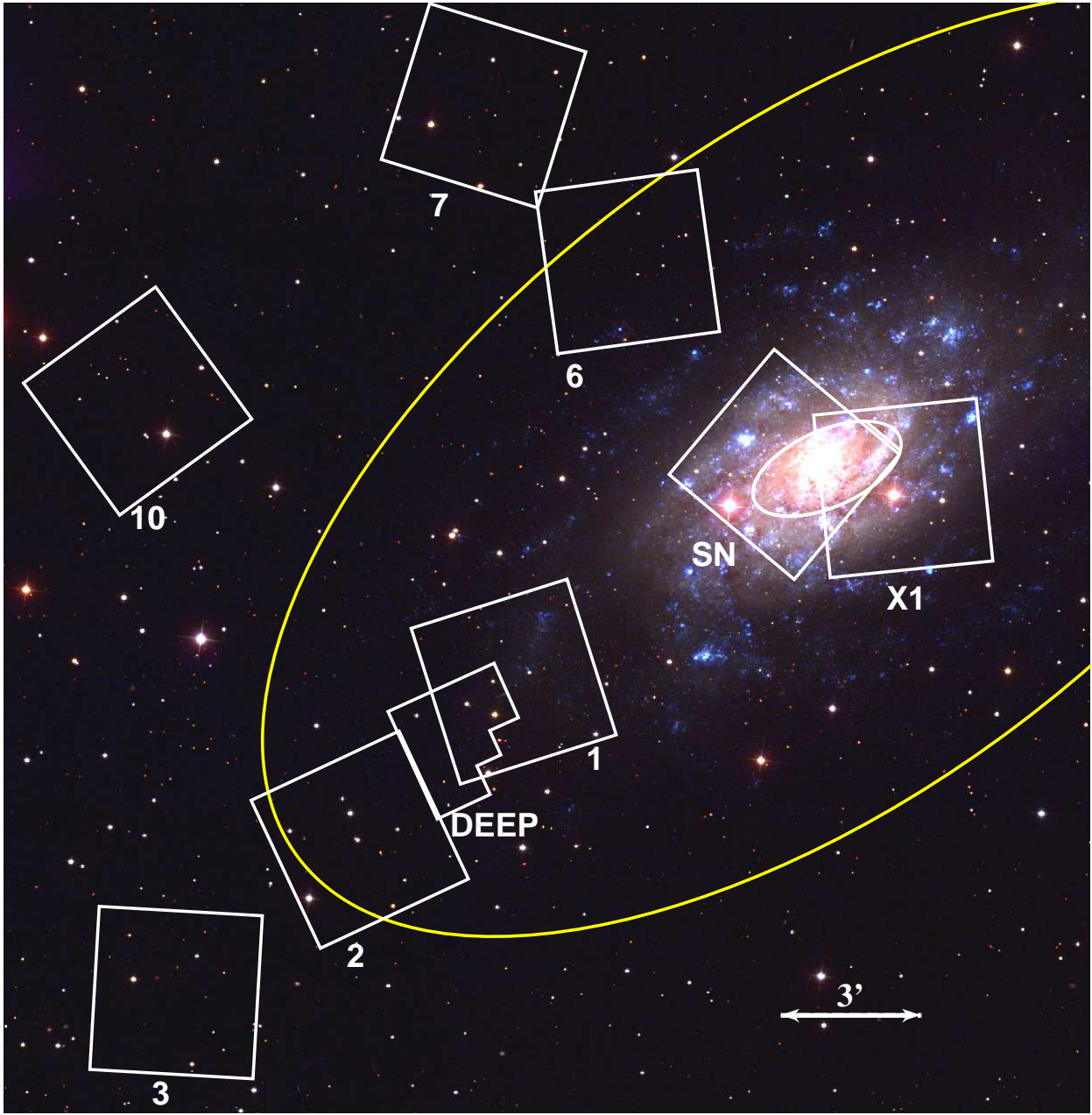,width=6.5in,angle=0}}
\caption{Footprints of our sample HST fields shown on a Sloan Digital
Sky Survey image of NGC~2403.  North is up and East is left. Fields are
labeled with shortened versions of their names given in
Table~\ref{obs_table}. Yellow ellipse marks the approximate B25 extent
of the galaxy \citep{nilson1973}.  White ellipse marks the division of
central fields described in the text and shown in
Figure~\ref{ellipse}.}
\label{image}
\end{figure}

We observed a field at large radius along the major axis of NGC~2403
as part of the ANGST survey from 2007-11-26 to 2007-12-01. The field
was observed with WFPC2 after the failure of ACS.  The telescope was
pointed to R.A.(J2000)=114.528667, decl.(J2000)= 65.504639 with PA\_V3
fixed at 69.995$^{\circ}$. The footprint of the field (DEEP) is shown
on a color image produced from Sloan Digital Sky Survey data in
Figure~\ref{image}.  To maximize the depth, full-orbit exposures were
obtained.  We took 12 such exposures through the F606W filter and 23
such exposures through the F814W filter.  These totaled 32400 s of
exposure in F606W and 62100 s of exposure in F814W.

We have also analyzed all relevant archival HST observations, which
included several GHOSTS (PI: de~Jong; PID: 11613;
\citealp{radburn2011}) fields along the major and minor axes, as well
as the 2 fields near the center of the galaxy (SN-NGC2403-PR, PID: 10182,
PI: Filippenko, \citealp{filippenko2004}; NGC2403-X1, PID: 10579, PI:
Roberts, \citealp{roberts2005}).  The exposure times, filters, and
depths of these data are given in Table~\ref{obs_table} (in order of
increasing distance from the galaxy center), and their
footprints are shown in Figure~\ref{image} (and labeled with a
shortened version of their name).

\subsection{Photometry}

\begin{figure}
\centerline{\psfig{file=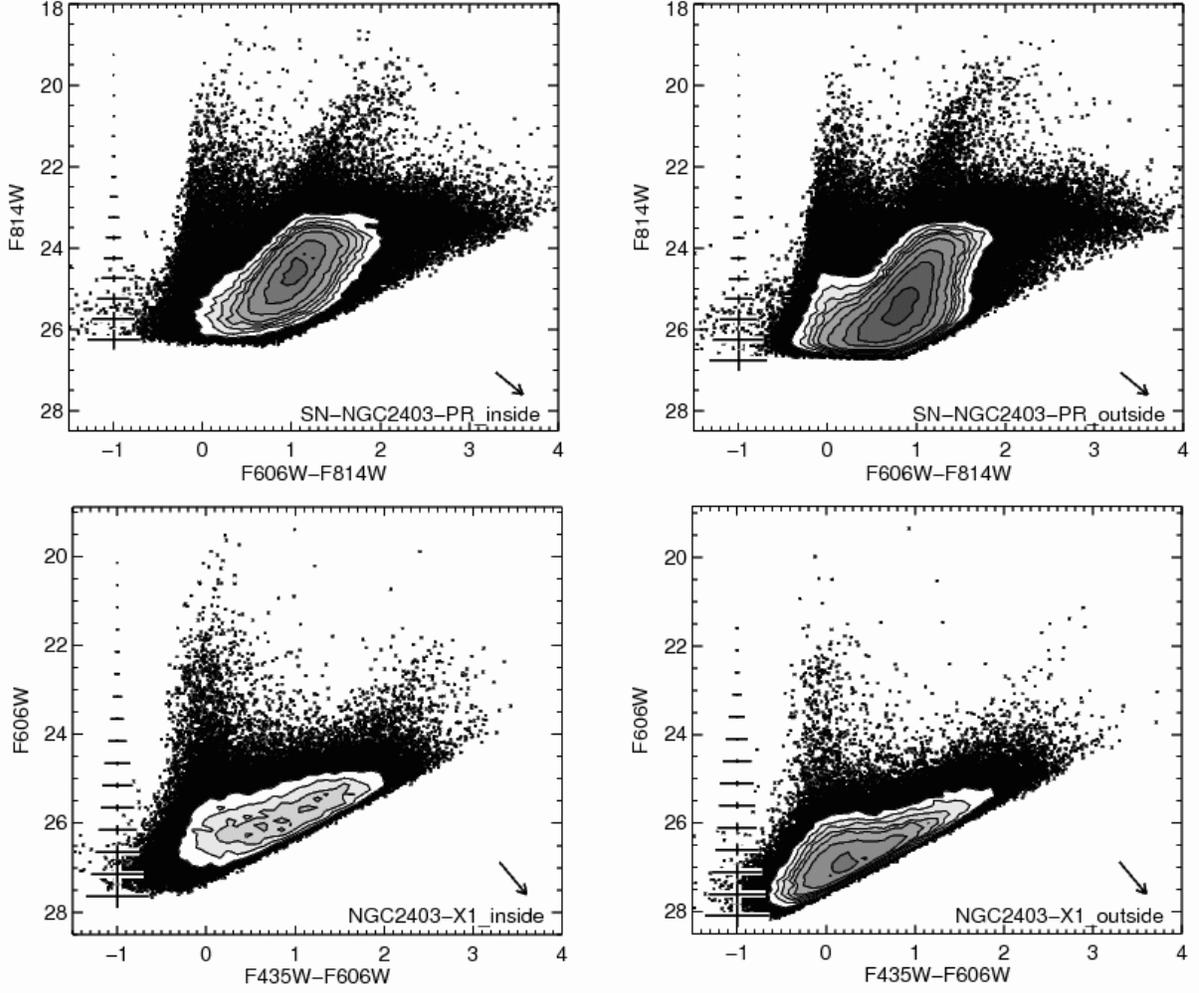,width=6.5in,angle=0}}
\caption{Color-Magnitude diagrams of the 2 inner HST fields used for
our study.  Photometry inside of the white ellipses in
Figures~\ref{image} and \ref{ellipse} is shown in the left column;
photometry outside is shown in the right column. Contours show the
density of points where the number of data points would otherwise
saturate the plot.  Error bars on the left side of each plot show the
mean photometric uncertainty, as calculated from counting statistics,
as a function of F814W magnitude.  Arrows in the lower right corner of
each plot show the direction of the reddening vector in the plotted
scale.}
\label{innercmds}
\end{figure}

\begin{figure}
\centerline{\psfig{file=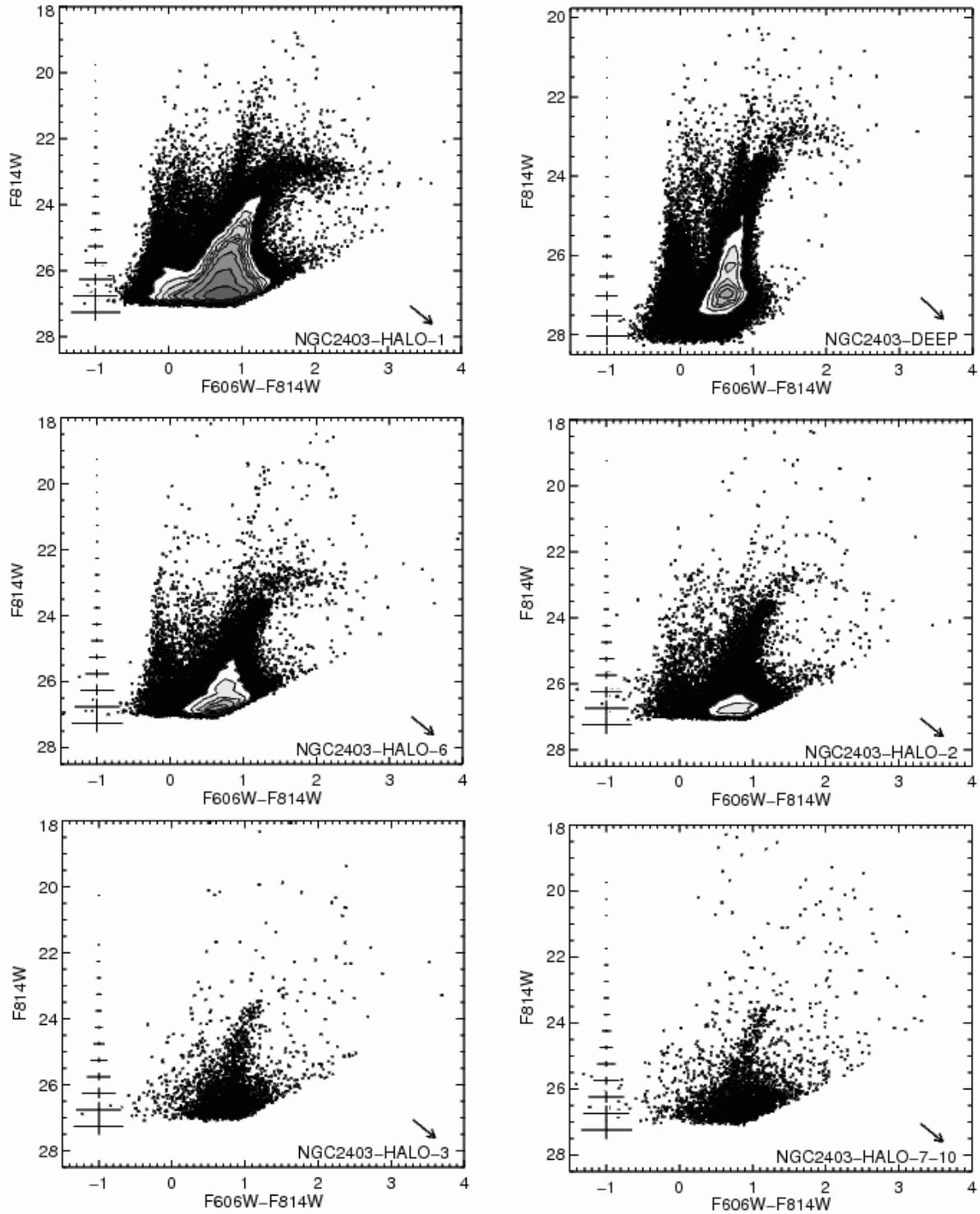,width=6.5in,angle=0}}
\caption{Color-Magnitude diagrams of the 7 outer HST fields used for
our study. From upper-left to lower-right, panels are ordered in
increased galactocentric radius. Fields 7 and 10 were combined into a
single CMD (lower-right) for the purposes of this figure as they are
both sparse and sample equivalent galactocentric radii.}
\label{outercmds}
\end{figure}

All photometry and artificial star tests were carried out as described
by the ANGST project paper \citep{dalcanton2009}.  Briefly, all WFPC2
photometry was derived with the package HSTphot \citep{dolphin2000}
and ACS photometry was measured using DOLPHOT, a general-purpose
stellar photometry package based on HSTphot that includes an
ACS-specific module.  Photometry was culled based on signal-to-noise
and the quality parameters of sharpness and crowding, as described in
\citet{dalcanton2009}.  The final color-magnitude diagrams (CMDs) for our
fields are shown in Figures~\ref{innercmds} and \ref{outercmds}.

We also ran several hundred thousand artificial star tests for each
field.  In each test, a fake star of known color and magnitude was
inserted into the data, and the photometry was rerun to attempt to
recover the photometry of the fake star.  These tests allowed us to
determine the recovery fraction and photometric errors as a function
of color and magnitude.  With this large number of fake stars, we
covered the full magnitude and color space of the real photometry,
including testing stars 1.5~mag below our limiting magnitude to
assess the likelihood of scattering faint stars into our sample.  Half
of the stars were randomly distributed in the relevant color-magnitude
space (the space covered by our real photometry), and the other half
followed the color-magnitude distribution of the observed stars.  This
sampling provided statistics for portions of the CMD where no stars
were observed as well as providing the best statistics for the areas
of the CMD where the most stars were observed.  

These tests showed that photometric errors in the uncrowded outer
fields were dominated by counting statistics, and stayed below 0.1 mag
brightward of $\sim$1 mag above the completeness limit, reaching a few
tenths of a mag by the completeness limit.  In the innermost regions
(Figure~\ref{innercmds}), our photometric errors were dominated by
crowding, and were therefore larger, reaching 0.1 mag $\sim$2 mag
above our completeness limit, and reaching $\sim$0.5 mag at the
completeness limit. The completeness limits given in
Table~\ref{obs_table} provide the magnitude where the recovered
fraction of artificial stars was 50\%.

\subsection{Modeling the Color-Magnitude Diagrams}

We fit the CMDs for each of the fields in our study using the software
package MATCH \citep{dolphin2002}.  The overall technique for deriving
the SFH using CMD fitting, as it has been applied for all ANGST
papers, is described in detail in \citet{williams2009}.  Output star
formation rates are renormalized to a \citet{kroupa2001} initial mass
function.  We derived the best-fitting mean extinction to the full
field with the distance fixed to the value found in
\citet{dalcanton2009}.  With these distance and extinction values
applied, we then run a series of 100 Monte Carlo (MC) tests.  In each
of these tests, a CMD is derived from the SFH that gave the best fit
to the data, including errors and completeness from the artificial
star tests.  This test CMD is then fitted to determine the precision
with which the input SFH was recovered.  The distribution of results
from these fits provides a reliable estimate of the uncertainty of our
SFH measurement given our data quality.

We have updated our method of estimating uncertainties compared to
\citet{williams2009}. To assess the systematic errors due to model
deficiencies, we performed fits to our MC realizations of the
best-fitting model solution with the models shifted in bolometric
magnitude and effective temperature \citep{dolphin2012}.  These shifts
account for the uncertainties due to any potential systematic offsets
between the data and models.  The uncertainties derived from these
tests for the inner fields represent the systematic uncertainties in
the dusty, highly crowded regions.  Those for the deep field represent
the systematics in the less crowded outer disk.

These systematic uncertainties due to deficiencies in the models will
affect all of our fits in the same manner, and therefore are relevant
only for comparisons between our measurements and those from other
techniques.  Because all of our data were analyzed homogeneously,
these systematic uncertainties are {\it not} relevant for comparisons
between different fields in our data set. To estimate the relative
uncertainty for differences between two fields analyzed using the same
models, we need only to assess the random errors due to the depth and
size of the stellar sample.  Therefore, for the ACS fields outside of
the central 2 fields, our MC error analysis did not include shifts
between the models and data, and the uncertainties show the
significance of relative differences between the observed fields.

\begin{figure}
\centerline{\psfig{file=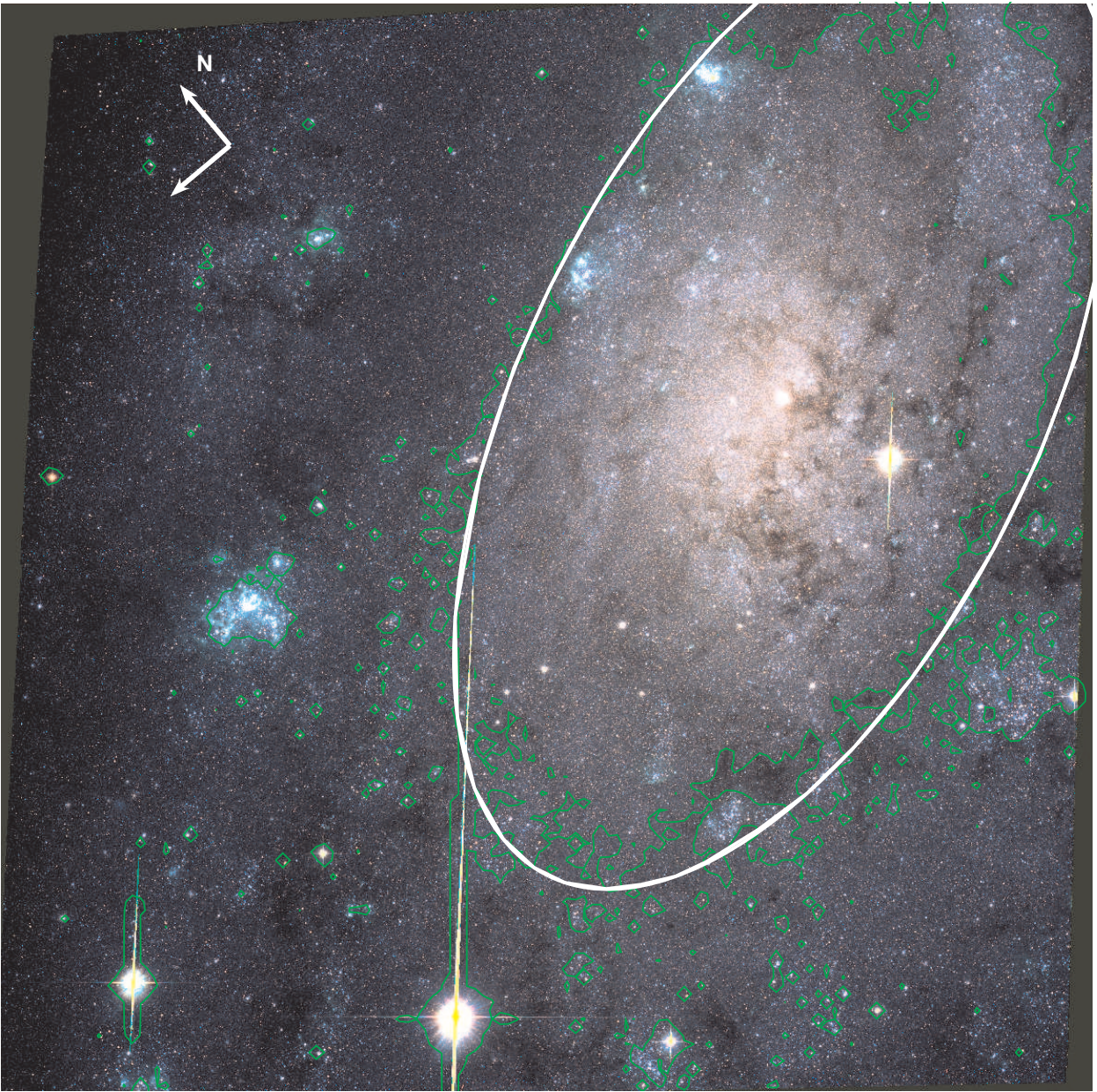,width=6.5in,angle=0}}
\caption{Our most central field (SN-NGC2403-PR).  Blue is F475W;
green, F606W; red, F814W.  The contour shows the F814W
%zeropoint = 25.53, flux is 0.13 cts/s/pixel
($\mu{\sim}21.2$ mag arcsec$^2$) isophote used to guide our field
division, shown as a white ellipse.}
\label{ellipse}
\end{figure}

When analyzing the densest, inner portions of the galaxy, we divided
both of the two innermost fields into 2 regions each: the most crowded
and dusty innermost region and the remainder of the field.  This
division is shown as the white ellipses in Figures~\ref{image} and
\ref{ellipse}, and corresponds roughly to an $I$-band surface
brightness of 21.2 mag arcsec$^{-2}$.  Separating the fields in this
way allowed us to isolate the portion of the galaxy most strongly
affected by crowding and differential reddening. Inside of this
ellipse, the median inclination-corrected galactocentric radii were
1.2 kpc for the SN field and 1.5 kpc for the X1 field; outside, they
were 2.6 kpc and 2.9 kpc for the two fields, respectively.

We found that photometry in the inner regions was 0.5 mag shallower
than in the outer regions of these fields, and shows clear evidence
for differential extinction. We therefore allowed for differential
extinction across the regions by introducing a spread in the model
colors and magnitudes along the reddening line.  The amplitude of this
spread was varied along a sequence of values ranging from 0 to 2
magnitudes.  The best fitting model out of this sequence was then
chosen.  For the central region, the best fitting value corresponded
to differential extinction of 1.6 magnitudes in both fields (i.e. the
stars were scattered along the reddening line by 1.6 mag).  The outer
region of the X1 field had a best-fitting value of 1.0 for the
differential extinction, and that of the SN field was 0.8.

Unfortunately, even including these effects in the models when fitting
these dusty and crowded regions, our MC uncertainty measurements
showed that we could not constrain the fraction of old stars, nor
their metallicity distribution, in the inner galaxy.  Therefore, to
fit the CMDs of these inner regions reliably for the young
populations, we assumed that their old stellar populations ($>$10~Gyr)
matched those of the DEEP field, with the SFRs for these ages simply
scaled by distance from the center assuming a scale length of 2.0 kpc.
This assumption is reasonable if the stars are well mixed on 10~Gyr
timescales.  We fixed the model for the old stars to these scaled
values only when fitting the CMDs of the different annuli of the inner
2 fields (X1 and SN).  By fixing the density and metallicity of old
stars in these innermost regions, we leveraged knowledge gained from
our deeper data to achieve a more reliable result for the total
populations from the dusty and crowded data.  Due to the large
differential reddening values, which substantially increased our
uncertainties, we used coarse age resolution to limit the SFR
uncertainties to less than 100\%. In the end, we used the fits to
these dusty and crowded inner regions of the disk only to constrain
the mean rate of star formation over the past 200 Myr.  Our results
for this quantity proved to be robust, as demonstrated by the
agreement for the common region inside of the F814W=21.2 isophote (see
Section \ref{recentsec}).

The outermost fields of the study (HALO-3,HALO-7,HALO-10) have only
sparsely populated red giant branches, which did not provide reliable
fits when allowed too many free parameters for the available data.
Therefore, in these outer-fields, we constrained the explored
metallicity to [Fe/H]${<}{-}$0.5.  This limit is above the known
metallicity of the galaxy inside of 10~kpc and is well-above the
metallicity measured for better-sampled fields outside of 10~kpc,
where no such boundary was implemented.  We found that with this limit
in place, the final fits provided the best reproduction of the
observed CMD features.

\section{Results}

\subsection{Cumulative Star Formation}

\begin{figure}
\centerline{\psfig{file=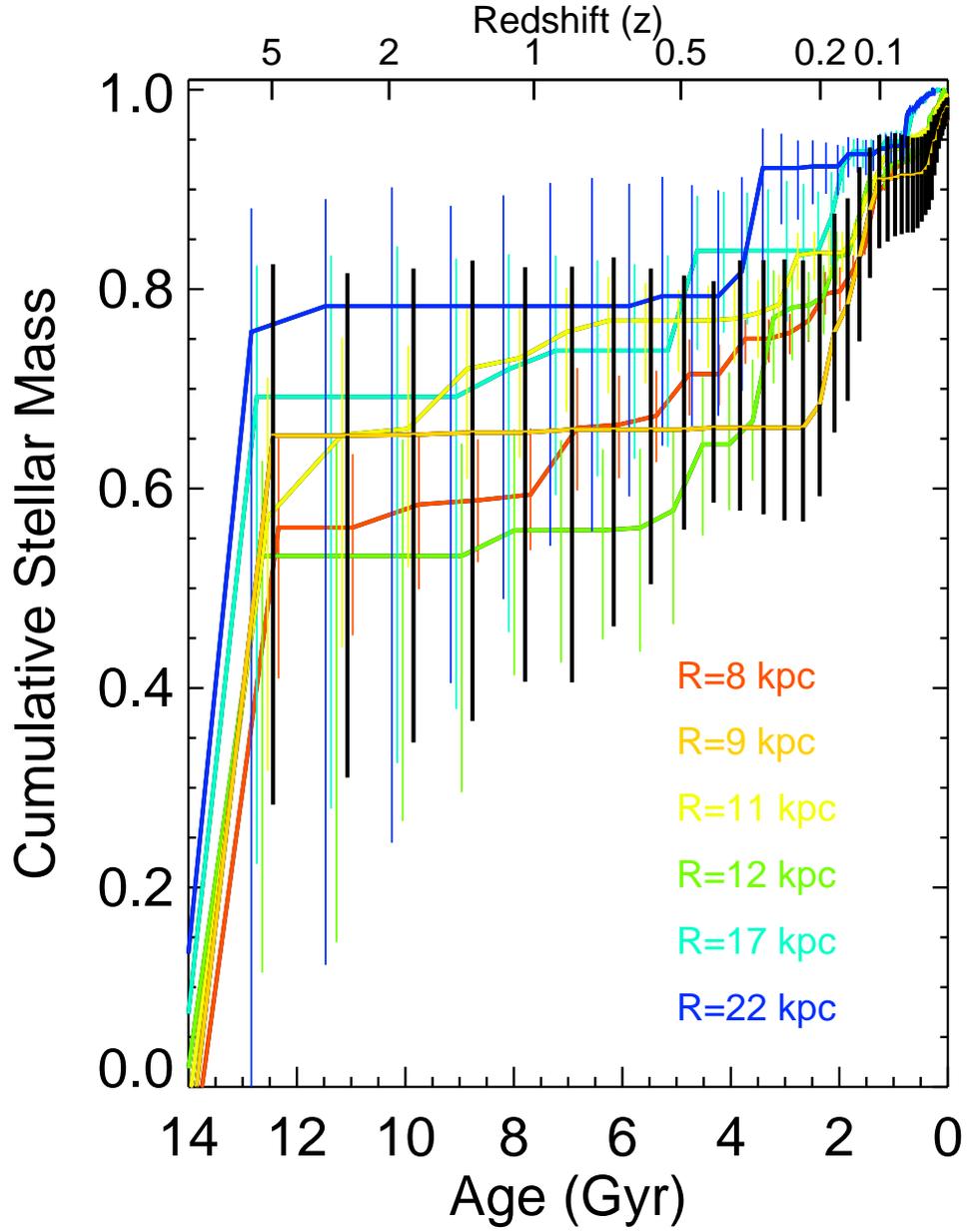,height=6.5in,angle=0}}
\caption{Cumulative SFHs for all of our measured regions, color-coded
by distance from the center, assuming all areas are dominated by the
disk population. The black uncertainties include systematic errors,
which would apply to all of the results together.  Thus, relative
comparisons should not include the systematic errors when comparing
the results to one another. All fields outside
of the inner disk show remarkably similar SFHs (consistent within
their relative uncertainties), suggesting little radial variation.}
\label{sfhs}
\end{figure}

In Figure~\ref{sfhs}, we show the cumulative stellar age distributions
of all of the fields sensitive to the old stellar populations
(i.e. all fields outside of 5~kpc except for the HALO-10 field, which
had too few stars to provide a reliable measurement). The different
fields are color-coded by distance from the galaxy center.  No trend
with radius is observed.

\begin{figure}
\centerline{\psfig{file=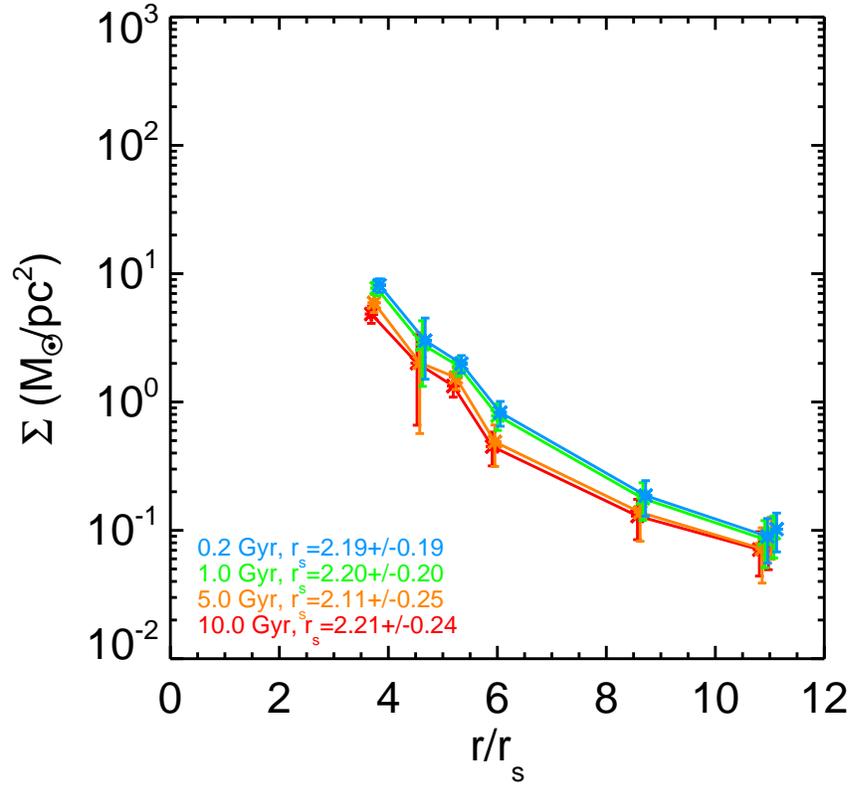,height=4.5in,angle=0}}
\caption{The radial surface density profile computed at several
epochs, showing little or no evolution in NGC~2403. }
\label{2403profile}
\end{figure}

Figure~\ref{2403profile} shows the surface density profile evolution
for the disk outside of 3 scale lengths, which is calculated from the
cumulative distributions from Figure~\ref{sfhs}.  No evolution in the
scale length of the surface density profile is seen, as expected given
that all of the cumulative distributions are consistent with one
another.  However, the profile does appear to flatten somewhat at
radii outside of $\sim$8 scale lengths.  This measurement is
consistent with the B12 star count analysis, which measured a more
extended structural component beginning at this radius.

Succinctly, our results confirm the result of B12 that the surface
density of the old stellar populations follows the exponential disk
profile out to $\sim$8 scale lengths.  However, here we also see that
stars of all ages are following a similar radial distribution,
including the youngest stars, which we discuss in detail below.

\subsection{Recent Star Formation Rate}\label{recentsec}

\begin{figure}
\centerline{\psfig{file=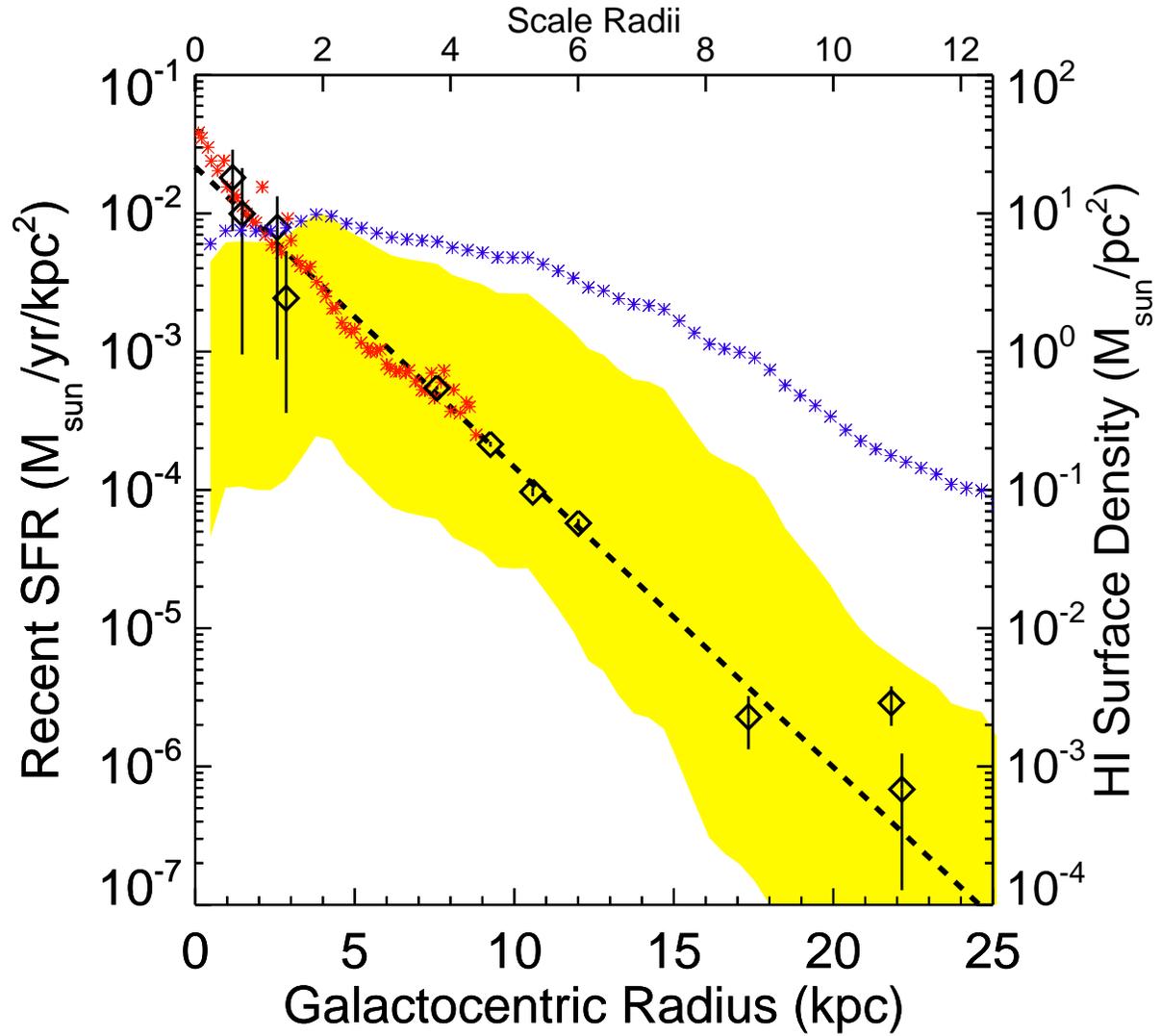,width=6.5in,angle=0}}
\caption{Recent SFR surface densities for all of our measured regions
as a function of deprojected galactocentric distance compared to those
measured in \citet{leroy2008} using Galex UV + Spitzer 24$\mu$ fluxes
(red asterisks). The H{\sc i}
surface density profile is also plotted (violet asterisks, right side
y-axis) along with the inferred SFR assuming the SFR-H{\sc i} surface
density correlation of \citet[][yellow shaded region]{bigiel2008}.}
\label{comp}
\end{figure}

In Figure~\ref{comp}, we show the mean SFR surface density over the
past 200 Myr as a function of radius.  The fractional uncertainties
for the inner fields are large due to the high crowding and
differential reddening, which is well-characterized by our MC tests.
We note the excellent agreement for the rates measured from both the
X1 and SN fields at these inner radii.  The uncertainties at
intermediate radii are small due to the lower crowding and
differential reddening, but still a large number of stars in the
sample.  At outer radii, the uncertainties increase again due to low
numbers of stars in the fields.

We overplot the combined Galex-Spitzer 24$\mu$-based SFR profile from
\citet{leroy2008}.  While our data probe SFR intensities a factor of a
hundred lower
($\sim$2$\times$10$^{-6}$~M$_{\odot}$~yr$^{-1}$~kpc$^{-2}$), the
agreement between the techniques is very good, as the SFRs from
integrated light follow the same exponential profile as recovered by
the resolved population measurements when averaged over the 200~Myr
timescale.  Furthermore, the normalization of the azimuthally-averaged
rates from \citet{leroy2008} agree with our rates, which were measured
for small regions at each radii.  These results suggests both that the
young stars are relatively well-mixed azimuthally in a 200 Myr
timescale and that the hybrid Galex-Spitzer 24$\mu$-based SFRs are
measuring an average rate covering a timescale of $\sim$200~Myr out to
at least 4 scale lengths.

The dashed line in Figure~\ref{comp} is the well-known exponential
disk profile of NGC~2403 as measured in $V$-band and RGB star counts
(\citealp[$h_r$=2~kpc][]{cepa1988}; B12).  This line is not a fit to
the data, other than a simple normalization.  Not only is the recent
star formation significantly detected at all radii, it also follows
the structure of the global disk for at least 11 scale lengths,
corresponding to nearly 4 orders of magnitude of star formation rate
surface density.

The unbroken exponential profile is again consistent with the lack of
a disk break and the lack of age gradients in the long-term
SFHs. Furthermore, it confirms the result of \citet{davidge2007} that
the radial distribution of the young stars traces that of the old
population, but extends that earlier result beyond 7 scale lengths to
11 scale lengths.  It is also consistent with the BIMA SONG results of
\citet{wong2002,wong2004}, which found that radial SFR profiles
followed the disk surface density, suggesting that pressure plays the
dominant role in the formation of molecular gas (and hence, the
formation of stars).  Thus, our measurements suggest that their
relations may hold to much lower SFRs than available for their study.

Finally, in Figure~\ref{comp} we overplot in blue the H{\sc i} profile
from the THINGS data \citep{leroy2008}, and shaded in yellow are the
SFRs that come from applying the relation of \citet{bigiel2008} to the
H{\sc i} profile.  The relative consistency supports the most recent
work on the relationship between gas surface density and SFRs.  No
threshold surface density is required for star formation, but there is
a decrease in star formation efficiency at low gas surface densities
\citep{bigiel2008,bigiel2010}.

\subsection{Age and Metallicity Gradients}\label{gradients}

\begin{figure}
\centerline{\psfig{file=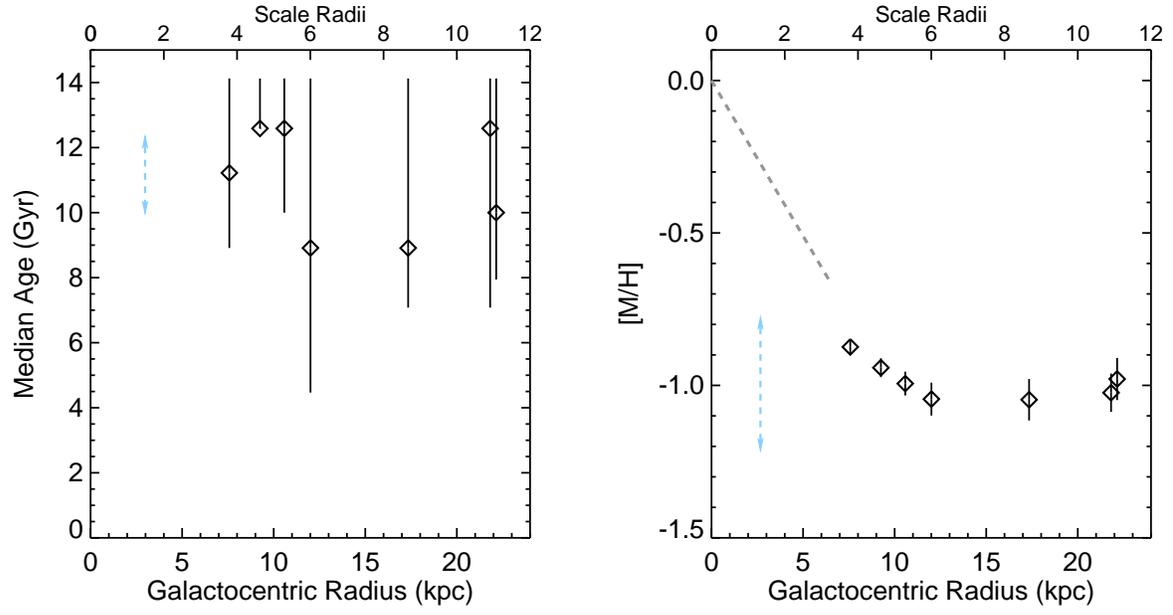,width=6.5in,angle=0}}
\caption{The median age (left) and mean metallicity (for stellar ages
$>$1~Gyr; right) %, and peak of a Gaussian fit to the RGB color
%distribution (23.6$<$F814W$<$24.6; right) 
of the stellar populations of NGC~2403 as a function of deprojected
galactocentric radius.  Dashed arrows show the systematic uncertainty
that could shift all points up and down in age by $\sim$2.5~Gyr or in
metallicity by $\sim$0.5~dex.  Dashed line shows the gas-phase
gradient from \citet{garnett1997}.}
\label{radial}
\end{figure}

In Figure~\ref{radial}, we plot the median ages and mean metallicities
as a function of radius (assuming all stars are in an inclined disk).
Inside of 5~kpc, we hand-fixed the metallicity and surface density of
the oldest stars, and therefore do not include those radii in these
plots.  The median ages are given at the age where the cumulative
stellar mass was closest to 50\%. The age uncertainties include all
time bins when the cumulative stellar mass was consistent with being
50\%.  These uncertainties therefore provide the range of median ages
consistent with our data, as inferred from our MC tests.  The absence
of a lower age error bar for our DEEP field relates to the best
constraint for old stars.  The age is restricted to the oldest age
bin, but the cumulative stellar mass was closer to 50\% at the end of
the time bin ($\sim$65\%) than at the beginning (0\%).  Plotted
metallicity errors are uncertainties in the mean, and do not reflect
the potential spread present in each field.  This metallicity spread
was typically $\sim$0.3 dex in each outer disk field.  The old
population dominates at all radii, and the mean metallicity shows
evidence for a negative gradient out to 12~kpc, then a constant value
out to large radii.

To first order, the lack of both an age gradient and an outer disk
metallicity gradient is consistent with the conclusions of B12.
Outside of 3 scale lengths, our data limit any age gradient to
be older than 8~Gyr.  Thus, if the median age of the disk at 9 kpc is
13~Gyr, as measured by our DEEP field, our data only allow a downward
age gradient to a minimum median age of 8~Gyr at 22~kpc.  Thus, the
steepest gradient allowed would be 0.4 Gyr kpc$^{-1}$.

We do see evidence for a metallicity gradient inside of 12~kpc (6
scale lengths) in the old stars ($>$1~Gyr) that dominate the stellar
population. The mean metallicity appears to decrease from
[Fe/H]$\sim{-}$0.8 at 4 scale lengths (8~kpc) to [Fe/H]$\sim{-}$1 at
$\sim$6 scale lengths (12~kpc).  This gradient ($-$0.05 dex
kpc$^{-1}$) is less steep than the $-$0.1 dex kpc$^{-1}$ gradient
measured for the gas phase metallicity out to 3 scale lengths (6~kpc)
by \citet{garnett1997}. The gradient out to $\sim$15~kpc is consistent
with the B12 measurement of a longer scale length for the metal-poor
RGB than the metal-rich RGB from 9--17~kpc.

Outside of 6 scale lengths, the metallicity gradient flattens,
remaining at [Fe/H]$\sim{-}$1 out to at least 11 scale lengths
(22~kpc). By mapping the RGB color to 10~Gyr isochrones, B12 estimated
a constant [Fe/H]$={-}1.0{\pm}0.3$ from 20--30~kpc, assuming that all
stars are older than 10~Gyr.  Thus, our value is consistent theirs,
but shows that the mean metallicity is flat in these regions to a
precision of 0.1 dex.

The flattening of the metallicity gradient appears correlated with the
surface density upturn in the RGB stars beginning at 8 scale lengths
seen by B12.  As they suggest, the upturn appears uncorrelated with
the star forming disk.  We clearly show in Figure~\ref{comp} that no
such upturn is seen in the SFR density.  The surface density upturn
appears due to a separate population that is uniformly very old and
well-mixed, leading to a constant metallicity with radius.

Previous studies have noted a metallicity gradient in the interstellar
medium and young stars in the inner $\sim$8~kpc.
\citep[e.g.,][]{garnett1997,davidge2007}.  We are not particularly
sensitive to metallicity gradients in younger stars, as there are few
CMD features that are sensitive to metallicity at ages younger than a
Gyr.  The metallicities in Figure~\ref{radial} are for the stars with
the ages $>$1~Gyr.  Furthermore, without metallicity measurements
inside of 3 scale lengths we cannot reliably address the possibility
of a metallicity gradient in the inner disk, even for the old stars.

Overall, our measured median ages are consistent with being the same
at all radii outside of 7~kpc, although the uncertainties are
relatively large.  However, our metallicity measurements show a
constant mean metallicity in the outer disk of NGC~2403 (to within 0.1
dex), spatially coincident with the surface density upturn in the old
population.  Therefore, the enhanced surface density appears to be due
to a population of very well-mixed old stars with relatively low
metallicity.

%\section{Discussion}
\section{Comparison with M~33 and NGC~300}

Our analysis of NGC~2403 provides some new information about the
physical properties of this very extended and apparently undisturbed
disk.  We now discuss how these properties compare with two other
bulgeless systems that can be studied in similar detail, M~33 and
NGC~300.  In particular, we describe how these few galaxies can help
shed light on the evolution of galaxy disks.

%\section{Comparison with M~33 and NGC~300}

\begin{figure}
\centerline{\psfig{file=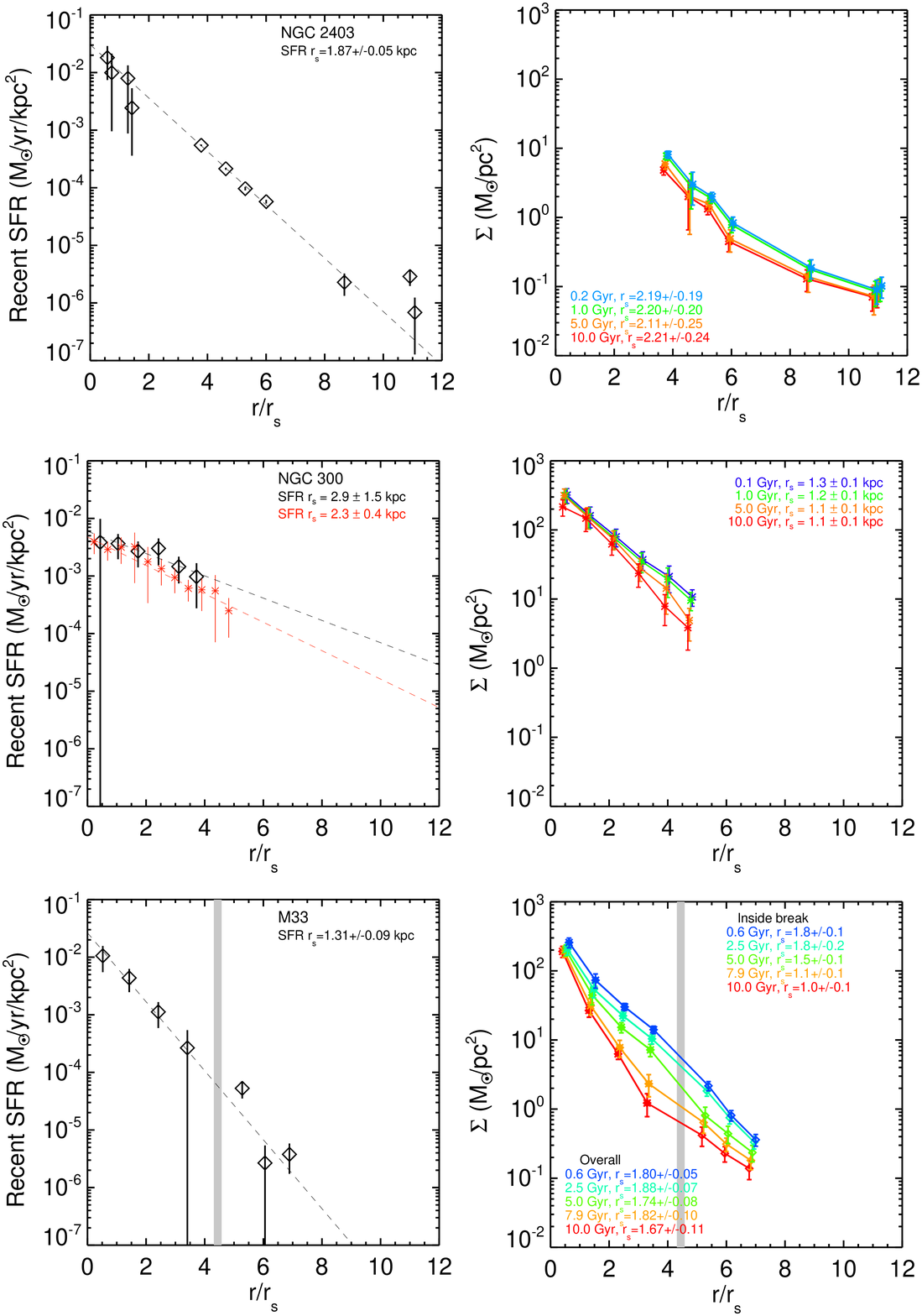,height=6.5in,angle=0}}
\caption{Comparison of the 3 Local Volume pure-disk galaxies. {\it
Left:} The recent star formation rate as a function of radius.  For
NGC~300, we augmented the data from \citet[][black
diamonds]{gogarten2010} with rates measured from Galex FUV + Spitzer
24$\mu$ imaging (red asterisks), using the method described in
\citet{leroy2008}. {\it Right:} The radial surface density profile
computed at several epochs, showing little or no evolution in NGC~2403
or NGC~300 compared to M~33. Dashed line for NGC~2403 is same as in
Figure~\ref{comp}.  Dashed lines for M~33 and NGC~300 are the best-fit
exponential to the star formation rate data points.  Thick vertical
gray line marks the radius of the M~33 disk break. Scale lengths used
for the X-axis are those in Table~\ref{summary}.  }
\label{compprofiles}
\end{figure}

In Figure~\ref{compprofiles}, we provide summary comparison plots of
NGC~2403 (this work), M~33 \citep{williams2009b}, and NGC~300
\citep{gogarten2010}, showing the recent SFR as a
function of radius, and the evolution of the disk surface density as
measured from the integral of the SFHs in radial bins.

The SFR profiles are shown as a function of disk scale length (as
measured from disk surface brightness), out to 11 scale lengths.
Therefore differences in the steepness of the profiles are showing
differences in how the star formation falls off relative to the
overall disk surface brightness. To improve our radial coverage, we
attempted to extend the star formation profile for NGC~300 by
including the {\it Galex} and {\it Spitzer} data to measure the mean
SFR out to a radius about 1 scale length larger than the available HST
data.  No break is seen.  We show the gradient extrapolated to 11
scale lengths, in Figure~\ref{compprofiles}.  A measurement of the SFR
of the far outer disk of NGC~300 would be of great interest to
complete this comparison.

As discussed earlier, the SFR in NGC~2403 follows the same
scale length as its overall disk surface brightness. Therefore, the
slope of the line in the top-left panel of Figure~\ref{compprofiles}
shows the SFR following the surface brightness profile.  The slope in
the NGC~300 panel is shallower, showing that the SFR of NGC~300 falls
off more slowly with radius than its overall disk surface brightness.
The same comparison for M33 shows that the its SFR falls very steeply
compared to its overall disk surface brightness.  Thus, the galaxy in
the densest environment (M~33) is the only disk of the three that
exhibits centrally-concentrated star formation.  Furthermore, M~33 is
the only disk of the three that shows a break in its surface
brightness profile \citep[B12;][]{bland2005,ferguson2007}.  These
results point to a possibility that environment is playing a role in
concentrating star formation and/or producing disk breaks.

These possible correlations with environment also carry through to the
evolution of structure of the galaxies' disks.  In the right side
panels of Figure~\ref{compprofiles}, the radial profiles of the three
galaxies at several different epochs (as calculated from our SFH
results) are shown.  Again, M~33 looks distinct from NGC~300 and
NGC~2403 systems.  M~33 has had a factor of 2 increase in its scale
length over the past 10~Gyr, whereas NGC~2403 and NGC~300 have both
had very little evolution in their disk scale lengths.  This scale
length evolution is the natural byproduct of M~33's radial age
gradient. Within the M~33 disk break ($\sim$8~kpc), the cumulative age
distribution shifts to younger ages as the radius increases
\citep{williams2009b}.  Outside of this break radius, the cumulative
age distribution shifts back to older ages \citep{barker2007} with
increasing radius.  

%\subsection{Disk Breaks in Bulgeless Galaxies}

Succinctly, in this very limited sample of three very nearby bulgeless
disk galaxies, we see strong similarities between the two more
isolated galaxies, NGC~300 and NGC~2403.  Both have little disk
evolution, star formation that is not concentrated relative to the
disk surface density, and no disk break. In contrast, M~33, which has
2 massive neighbors, shows clear evidence of disk evolution, a steep
fall-off in SFR per scale length, and a disk break.  There is clearly
more work needed to understand the production and evolution of disk
breaks, and NGC~2403 is an important object to understand in this
context.  In particular, the structure of the NGC~2403 disk requires
that any model of disk break formation must also be able to produce an
{\it unbroken} star forming disk out to very low gas densities
($\lap$10$^{-1}$~M$_{\odot}$~pc$^{-2}$).

\section{Conclusions}

We have performed resolved stellar photometry on 9 HST fields in
NGC~2403 covering 11 scale lengths of the disk.  We fit the CMDs of
the stars in these regions to determine their age distributions and
mean metallicities.  We compared our results with those of previous
studies to provide a picture of the evolution of NGC~2403.  To put
this evolutionary history in context, we compared it to those of its
nearby morphological twins M~33 and NGC~300.

Our results suggest little radial variation in the age distribution of
the stellar populations in NGC~2403 with galactocentric distance,
consistent with previous work from star counts.  Furthermore, we find
that the surface density of star formation in the disk follows the
same exponential profile as the overall surface brightness profile of
the disk out to 11 scale lengths, covering 4 orders of magnitude of
intensity.

In addition to these strong indications of the undisturbed nature of
the NGC~2403 disk, the H{\sc i} rotation curve is known to be
extremely regular \citep{deblok2008}, showing no warps.  Indeed, it
has perhaps the most symmetric velocity field in the THINGS sample.
All of these attributes point strongly to NGC~2403 having had no
previous strong gravitational encounters. Our results therefore
suggest that NGC~2403 is falling into the M~81 group for the first time,
as any significant interaction with the group would almost certainly
have disturbed the very low density far outer disk, which we see as
very much intact.

Unlike the star formation and H{\sc i} surface density, the
metallicity of NGC~2403 may not follow a simple gradient.  The
metallicity decreases out to a radius of at least 12~kpc, then levels
off at some point between 12 and 18~kpc.  We note that B12 found that
the surface brightness profile of the old stars in NGC~2403 flattens
beyond 8 scale lengths. They attribute this very extended profile to a
separate galaxy component, unrelated to the star-forming disk.  The
uniform metallicity of this component with radius suggests that it is
very old and well-mixed. Further interpretation of these properties
will await deeper, wide-field data.

Our NGC~2403 results are very similar to the results for the
NGC~300 disk, suggesting comparable paths of evolution.  On the other
hand, M~33 has more centrally-concentrated star formation, a clear disk
break, and a strong age gradient, potentially as a result of its more
crowded environment in the Local Group.  The lack of a break or age
gradient in NGC~2403 shows that disk formation theory
must provide a means for producing an unbroken disk of both young and
old stars (with similar scale lengths) to
very low surface densities in an isolated environment, as we have
observed in NGC~2403.

Finally, a detailed study of the NGC~300 stellar disk out to large
radii would provide an interesting comparison, as NGC~300 does have a
warped H{\sc i} disk.  This warp may be due to the proximity of
NGC~55, which puts NGC~300 in an environment that lies between that of
the complete isolation of NGC~2403 and the massive companions of M~33.
Unfortunately, comparable HST data on the outer disk of NGC~300 are
not currently available.  The detailed studies of M~33 out to large
radii clearly indicate a different structure and history compared to
NGC~2403, perhaps due in large part to its proximity to and previous
interaction with M~31.  Such a study of the outer disk of NGC~300 will
provide a complete picture of all 3 local pure disk galaxies.

Support for this work was provided by NASA through grants GO-10915 and
GO-11986 from the Space Telescope Science Institute, which is operated
by the Association of Universities for Research in Astronomy,
Incorporated, under NASA contract NAS5-26555. We acknowledge the anonymous
referee's attention to details of our writing style and presentation.

%\bibliography{apjmnemonic,references}
%\bibliographystyle{apj}

\clearpage

 \begin{deluxetable}{ccccccccc}
\tablewidth{16.5cm}
\tablecaption{Summary of Galaxy Properties}
\tabletypesize{\footnotesize}
\tablehead{
\colhead{Galaxy} &
\colhead{Dist. (Mpc)\tablenotemark{a}} &
\colhead{Morphology\tablenotemark{b}} &
\colhead{$r_s$ (kpc)\tablenotemark{c}} &
\colhead{M$_{\rm V}$\tablenotemark{b}}&
\colhead{M$_{\rm K}$\tablenotemark{d}} &
\colhead{M$_{\rm HI}$}\tablenotemark{b} &
\colhead{[Fe/H]\tablenotemark{e}} &
\colhead{SFR (M$_{\odot}$ yr$^{-1}$)\tablenotemark{f}}
}
\startdata
%M~33 & 0.8 & SA(s)cd & -19.4 & 4.1 (-20.4) & 7.2 (-17.3) & 0.02\\
%NGC~300 & 2.0 & SA(s)d & -18.5 & 6.4 (-20.1) & 9.2 (-17.3) & 0.02\\
%NGC~2403 & 3.2 & SAB(s)cd & -19.5 & 6.2 (-21.3) & 9.6 (-17.9) & 0.02\\
M~33 & 0.8 & SA(s)cd & 1.8 & -19.4 & -20.4 & -17.3 & -0.8 & 0.03\\
NGC~300 & 2.0 & SA(s)d & 1.3 & -18.5 & -20.1 & -17.3 & -0.3 & 0.01\\
NGC~2403 & 3.2 & SAB(s)cd & 2.0 & -19.5 & -21.3 & -17.9 & -0.7 & 0.03\\
\enddata
\tablenotetext{a}{M~33 distance from \citet{lee2002}; others from
  \citet{dalcanton2009}}
\tablenotetext{b}{Morphologies, V-band and H{\sc i} absolute magnitudes were calculated applying the distance modulus (see table-note a) and dust extinction from \citet{schlegel1998} to the apparent magnitudes from \citet{devaucouleurs1991}.}
\tablenotetext{c}{Disk scale length (\citealp{williams2009b,gogarten2010}; B12)}
\tablenotetext{d}{K-band absolute magnitudes are based on the K$_{\rm tot}$ magnitudes of \citet{jarrett2003}}
\tablenotetext{e}{\citealp{kim2002,kudritzki2008}; B12}
\tablenotetext{e}{Star formation rates are the integrals of the SFR
  profiles (assuming azimuthal symmetry) measured here (see Figure~\ref{compprofiles}).}
\label{summary}
\end{deluxetable}

\begin{deluxetable}{cccccccc}
\tablewidth{16.5cm}
\tablecaption{Summary of Data and Photometry Measurements}
\tabletypesize{\footnotesize}
\tablehead{
\colhead{Proposal} &
\colhead{Target} &
\colhead{Camera} &
\colhead{Filter} &
\colhead{Exposure (s)}&
\colhead{Stars} &
\colhead{Radius (kpc)} &
\colhead{$m_{50\%}$} 
}
\startdata
10182 & SN-NGC2403-PR & ACS & F606W &    700 & 405516 & 2.0 & 26.18\\
10182 & SN-NGC2403-PR & ACS & F814W &    700 & 405516 &  2.0 & 25.54\\
10579 & NGC2403-X1 & ACS & F435W &   1248 & 154761 & 2.4 & 26.91\\
10579 & NGC2403-X1 & ACS & F606W &   1248 & 154761 & 2.4 & 26.44\\
10523 & NGC2403-HALO-1 & ACS & F606W &    710 & 101951 & 7.6 & 27.65\\
10523 & NGC2403-HALO-1 & ACS & F814W &    710 & 101951 & 7.6 &  26.80\\
10915 & NGC2403-DEEP & WFPC2 & F606W &  32400 & 30617 & 9.3 & 28.05\\
10915 & NGC2403-DEEP & WFPC2 & F814W &  62100 & 30617 & 9.3 & 27.20\\
10523 & NGC2403-HALO-6 & ACS & F606W &    720 & 25350 & 10.6 & 27.60\\
10523 & NGC2403-HALO-6 & ACS & F814W &    720 & 25350 & 10.6 & 26.92\\
10523 & NGC2403-HALO-2 & ACS & F606W &    740 & 15619 & 12.0 & 27.70\\
10523 & NGC2403-HALO-2 & ACS & F814W &    745 & 15619 & 12.0 & 26.93\\
10523 & NGC2403-HALO-3 & ACS & F606W &    735 & 3102 & 17.3 & 27.70\\
10523 & NGC2403-HALO-3 & ACS & F814W &    735 & 3102 & 17.3 & 26.92\\
10523 & NGC2403-HALO-7 & ACS & F606W &    745 & 2275 & 21.8 & 27.70\\
10523 & NGC2403-HALO-7 & ACS & F814W &    745 & 2275 & 21.8 & 26.92\\
10523 & NGC2403-HALO-10 & ACS & F606W &    730 & 1627 & 22.1 & 27.70\\
10523 & NGC2403-HALO-10 & ACS & F814W &    730 & 1627 & 22.1 & 26.81\\
\enddata
\label{obs_table}
\end{deluxetable}

\end{document}